\documentclass[sigconf]{acmart}

\AtBeginDocument{%
  \providecommand\BibTeX{{%
    \normalfont B\kern-0.5em{\scshape i\kern-0.25em b}\kern-0.8em\TeX}}}

\copyrightyear{2025}
\acmYear{2025}
\setcopyright{acmlicensed}\acmConference[CHI '25]{CHI Conference on Human Factors in Computing Systems}{April 26-May 1, 2025}{Yokohama, Japan}
\acmBooktitle{CHI Conference on Human Factors in Computing Systems (CHI '25), April 26-May 1, 2025, Yokohama, Japan}
\acmDOI{10.1145/3706598.3713431}
\acmISBN{979-8-4007-1394-1/25/04}

\newcommand{\fac}{\textsc{AutoCopilot}}%
\newcommand{\sac}{\textsc{GuidedCopilot}}%
\newcommand{\sav}{\textsc{GuidedCopilotVisual}}%
\newcommand{\saa}{\textsc{GuidedCopilotADP}}%

\usepackage{tikz}
\usepackage{amssymb}
\usepackage{xcolor}
\usepackage{multirow}
\usepackage{makecell}
\usepackage{amsmath}
\newcommand{\change}[1]{{\textcolor{black}{#1}}}
\begin{document}


\title[Do It For Me vs. Do It With Me]{Do It For Me vs. Do It With Me: Investigating User Perceptions of Different \change{Paradigms} of Automation in Copilots for Feature-Rich Software}

\renewcommand{\shortauthors}{}
\author{Anjali Khurana}
\orcid{0000-0002-2730-5512}
\affiliation{%
  \institution{Computing Science \\ Simon Fraser University}
   \city{Burnaby}
  \state{BC}
  \country{Canada}
  }
  \email{anjali\_khurana@sfu.ca}

\author{Xiaotian Su}
\orcid{0009-0004-0548-1576}
\affiliation{%
  \institution{Computer Science \\ ETH Zürich}
  \city{Zürich}
  \state{}
  \country{Switzerland}
  }
  \email{xiaotian.su@inf.ethz.ch}

\author{April Yi Wang}
\orcid{0000-0001-8724-4662}
\affiliation{%
  \institution{Computer Science \\ ETH Zürich}
  \city{Zürich}
  \state{}
  \country{Switzerland}
  }
  \email{april.wang@inf.ethz.ch}
  
\author{Parmit K Chilana}
\orcid{0009-0007-0173-1752}
\affiliation{%
 \institution{Computing Science \\ Simon Fraser University}
   \city{Burnaby}
  \state{BC}
  \country{Canada}
  }
  \email{pchilana@cs.sfu.ca}

\renewcommand{\shortauthors}{Khurana et al.}

\begin{abstract}
Large Language Model (LLM)-based in-application assistants, or \textit{copilots}, can automate software tasks, but users often prefer \textit{learning by doing}, raising questions about the optimal level of automation for an effective user experience. We investigated \change{two automation paradigms by designing and implementing} a fully automated copilot (\fac{}) and a semi-automated copilot (\sac{}) that automates trivial steps while offering step-by-step visual guidance. In a user study (N=20) across data analysis and visual design tasks, \sac{} outperformed \fac{} in user control, software utility, and learnability, especially for exploratory and creative tasks, while \fac{} saved time for simpler visual tasks. A follow-up design exploration (N=10) enhanced \sac{} with task-and state-aware features, including in-context preview clips and adaptive instructions. Our findings highlight the \change{critical role of user control and tailored guidance in designing the next generation of copilots that enhance productivity, support diverse skill levels, and foster deeper software engagement}. 

\end{abstract}







\maketitle

\section{Introduction}

\begin{figure*}[!t]%
\centering
    {\includegraphics[width=0.95\linewidth]{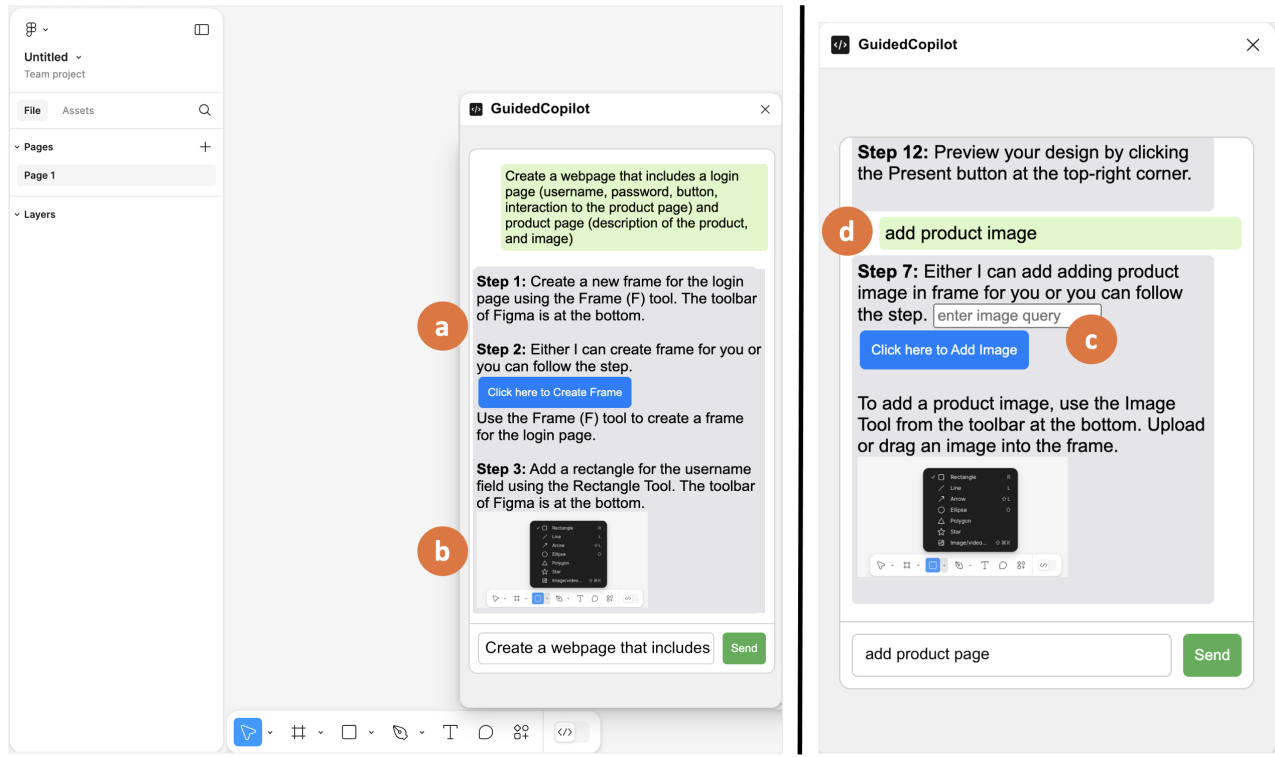} }%
      \vspace{-7pt}
    \caption{\sac{}, a novel semi-automatic copilot: (a) Copilot assistance is structured to provide step-by-step guidance along with semi-automation for only repetitive or trivial steps in the task; (b) Visual references of the UI elements in-context to user's tasks and application are provided within the step-by-step guidance; (c) Users have control over editing the LLM extracted entities from their query before the semi-automation is performed; (d) Up-to-date mixed-medium follow-up responses are provided. To see the contrast with the fully automated copilot assistance, please see \fac{} in Figure \ref{FullAuto_ui}.}%
   \label{SemiAuto_ui}%
\end{figure*}

Human-Computer Interaction (HCI) research has long been exploring in-application AI-based assistants \cite{Pangoli,Toby_multimodal, Khurana_chatrex, Robinson_meyer_2015, Horvitz_2007, Allen} designed to assist users in navigating feature-rich software applications. With the advancements in Generative AI and pre-trained Large Language Models (LLMs) \cite{AI_2022, brown_llm, copilot, Adobe_2023}, a new generation of AI assistants is emerging, capable of automating a wide range of complex tasks. 
These LLM-powered assistants, often branded as \textit{copilots} \cite{cacm_article, copilot, Adobe_2023}, can offer targeted, flexible assistance based on users' natural language description of help needs \cite{Johnny}. \change{By integrating context-sensitive automation dynamically into user workflows \cite{copilot, cacm_article, bansal2024challenges},} \change{these modern copilots} such as Microsoft 365 Copilot \cite{copilot}, Adobe Firefly \cite{Adobe_2023}, and Figma AI \cite{figma} are exploring ways to fully automate software tasks, redefining user expectations of what generative AI can achieve and pushing the boundaries of productivity and creativity.

\change{Despite the promise of fully automated copilots, they introduce distinct challenges due to their open-ended nature, inherent complexities, and wide-ranging failure modes \cite{bansal2024challenges}}. Users often need to repeatedly review AI outputs and refine their inputs to align them with their intent, increasing overall effort \cite{ivie, businessinsiderCanceledMicrosoft, Grounded_Copilot, Vaithilingam}. \change{Addressing these challenges requires effective human-agent collaboration \cite{shneiderman1997direct, bansal2024challenges}, which involves providing appropriate information \cite{Amershi_guidelines, shneiderman1997direct} and understanding the dynamics of user interaction across diverse AI scenarios to avoid false assumptions \cite{cacm_article, liao2023ai}.} 
Fully automated copilots often also overlook a key principle from HCI and software learnability research: users often prefer to \textit{learn by doing} when \change{exploring software features \cite{Carroll}}, underscoring the need for balance between guidance and automation \cite{Khurana_iui}. Research also shows that \change{step-by-step instructions with GUI visuals \cite{Khurana_chatrex, Toby_multimodal, palmiter1991animated} and demonstrations \cite{Lafreniere_Tutorials, Novick_tutorials, Grossman_ToolClips} are essential for improving feature discovery, retention, and learning outcomes \cite{Chilana_lemonaid, Grossman_learnability}.}

In this work, we investigate how humans and copilots can collaborate within feature-rich software and what the concept of a “copilot” truly means to users \cite{cacm_article}. We examine \change{two distinct design paradigms for automation \cite{endsley2017here, Horvitz_2007, bainbridge1983ironies}:} whether users prefer copilots that fully automate tasks \change{(“Do It For Me”), as seen in fully automated systems, or those that serve} as assistants, combining semi-automation for repetitive tasks with guided support and instructions for learning complex tasks \change{(“Do It With Me”)}. We investigate how these \change{design paradigms} (semi-automation vs. full automation) impact task completion and user perceptions of software copilots, considering factors such as \change{user expertise (novice vs. expert), familiarity with LLMs, and the nature of the tasks} (fixed vs and creative, exploratory). 

To study these design paradigms, we designed and implemented two in-application copilot interventions. The first, 1) \fac{} \change{(“Do It For Me”)}, fully automates software tasks (Figure \ref{FullAuto_ui}) based on the user’s textual prompt, inspired by \change{state-of-the-art copilot assistants (e.g., \cite{copilot, Firefly, figma})}. The second, \sac{} \change{(“Do It With Me”)}, is a novel semi-automatic copilot that automates only trivial or repetitive tasks while offering step-by-step \change{visual guidance to help users locate UI elements} (Figure \ref{SemiAuto_ui}). \sac{} integrates visuals into its responses, allows users to \change{initiate the automation process} and enables corrections before proceeding. The research questions guiding this investigation were:\begin{itemize}
       \item \change{RQ1: How do users perceive the utility, control, and potential for software learnability of in-application copilots designed using two different paradigms: full automation (“\textit{Do It For Me}”) and semi-automation (“\textit{Do It With Me}”)?}
    \item RQ2: How can we design an \change{in-application semi-automated} copilot that provides semi-automation along with \change{step-by-step visual guidance}?
\end{itemize}

We conducted a within-subject controlled experiment (N=20) and follow-up interviews to compare the strengths and weaknesses of \fac{} vs. \sac{} embedded in two different software applications: \textit{Google Sheets}, an online spreadsheet application and \textit{Figma}, an online user interface design application. 
While \fac{} was valued by some participants, \change{particularly males with CS backgrounds, for saving time on specific tasks}, \sac{} consistently outperformed in user control, software utility, and software learnability. These findings were further supported by higher task completion rates and accuracy scores for \sac{}. Participants expressed that \sac{} provided higher user control when performing or customizing complex tasks, whereas \fac{}'s full automation approach often did not align well with task requirements, leading to time-consuming rounds of trial-and-error and debugging. 

Although participants appreciated \sac{}’s in-context assistance, some felt that it provided overly detailed instructions for already-familiar tasks, while others struggled to map visuals in the chat to the corresponding UI elements.  To address this, we explored design improvements to enhance semi-automatic copilots by integrating real-time context of users' progress and the application state. We developed two key features (See Figure \ref{SemiAuto_vst}): \sav{} (Visual Step-through), which embeds context-specific preview clips in the software interface; and \saa{} (Adaptive Mixed-medium), which adapts and tailors instructions based on user needs and the state of their task progress. In a follow-up usability study (N=10), expert users appreciated \saa{}'s ability to adapt instructions to their proficiency levels, while novices valued the clickable navigation and the ability to skip steps. All participants found the preview clips helpful for building a mental model of the user interface, as they directly highlighted the relevant icons needed to complete each step based on users' progress and application state.

The key contributions of this paper are threefold: (1) \change{we examine two distinct design paradigms for automation through} the design and implementation of \sac{}, a novel copilot that combines semi-automation with step-by-step \change{visual guidance}; and \fac{}, a fully automated copilot; (2) empirical insights into the strengths and weakness of \fac{} vs. \sac{}, for completing software tasks, focusing on perceptions of software utility, user control and potential for software learnability; and, (3) a further design exploration of \sac{}, introducing two new features, \sav{} and \saa{}, that can enhance visual integration and deliver more targeted, task- and state-aware assistance, pushing the boundaries of in-context copilot support. We synthesized the insights from our studies into key factors (See Figure \ref{dimesnion}, Section \ref{dimension_framework}) and levels of automation along with guidance to consider for effective human-AI collaboration in feature-rich software environments. These findings offer important implications for HCI and AI research by providing a clearer understanding of how to balance automation and user control in software copilots. Future research and applications can build on these insights to develop more adaptive and user-centered AI assistants that improve task efficiency, user autonomy, and software learning experiences.

\section{Related Work}
This research builds upon insights from prior work on software learnability, LLM-based in-application assistants, and challenges and opportunities in human-AI interaction in domains such as programming and software development.

\subsection{Learning and Seeking Help for Feature-Rich Software}

Seeking help for feature-rich software is often challenging due to scattered learning resources that require precise queries for successful retrieval \cite{Furnas, Kiani_help, Grossman_learnability}. Despite the expansion of help formats beyond traditional documentation and manuals \cite{Rettig, Novick_manual}, novice end-users face the \textit{vocabulary problem} \cite{Furnas} as they try to locate and apply relevant help from online tutorials, Q\&A or FAQ sites, blogs, dedicated forums \cite{Kiani_help}, and videos \cite{Kim_videos, Lafreniere_Tutorials}. Many resources \change{are either outdated or too general to address specific needs} \cite{Kiani_help, wang2018leveraging}. While online communities offer targeted and personalized help, they often involve delays and social barriers that discourage user participation \cite{Kiani_help, wang2018leveraging}. \change{In-context help techniques} \cite{Chilana_lemonaid, matejka2011ip, social_cheatsheet, Lafreniere_incontext, Hartmann, Brandt_joel, Grossman_ToolClips} that assist people within their current task and/or application \change{and step-by-step guidance with visuals and demonstrations} \cite{yeh2011creating, Lafreniere_Tutorials, Novick_tutorials, Grossman_ToolClips, Khurana_chatrex, Toby_multimodal, palmiter1991animated} can be particularly useful for helping users within the context of their tasks \cite{Khurana_chatrex, Kiani_help, Chilana_lemonaid, matejka2011ip}.

\change{AI-based in-application assistants have also been explored in HCI and AI research to provide customized}, context-specific solutions within software environments \cite{Alyssa_Glass_toward, ramachandran2005providing, myers2006answering, Justin_cranshaw2017calendar, Khurana_chatrex, Toby_multimodal}. Early systems, such as SmartAidè \cite{ramachandran2005providing} and Crystal \cite{myers2006answering}, employed Machine Learning (ML) and Natural Language Processing (NLP) to interpret user intent and offer guidance. \change{However, unlike modern copilots powered by generative AI foundation models capable of tackling complex, open-ended tasks,} the \change{traditional AI} systems \change {based on simpler models} \change{\cite{endsley2017here, heer2019agency, shneiderman2020human, Alyssa_Glass_toward, ramachandran2005providing, myers2006answering, Justin_cranshaw2017calendar, bansal2024challenges}} were \change{limited to handling} narrow range of application-specific queries. Users often found the text-based assistance provided by these systems challenging to locate and use within the application's interface \cite{Alyssa_Glass_toward, ramachandran2005providing, myers2006answering, Justin_cranshaw2017calendar}. Recent systems, such as Appinite \cite{Toby_multimodal} and ChatrEx \cite{Khurana_chatrex} \change{offer usability improvements by combining text} and visual cues for task-specific queries. However, these systems still face challenges with \change{training sets, limited adaptability across applications \cite{Toby_multimodal} and difficulties in supporting ambiguous or contextually complex} queries \cite{Khurana_chatrex}. Building on these principles and insights from prior work, our novel semi-automatic copilot, \sac{}, automates only trivial or repetitive tasks while providing step-by-step guidance with visual references. In our follow-up design exploration, we further explored variations of \sac{} providing assistance dynamically adapting to the user's progress and the state of the application.

\subsection{Emergence of In-Application Software Copilots and LLM-based Assistants}
 LLM-based assistants have emerged as a “one-stop solution” \cite{AI_2022, prompt_catalogue}, enabling users to seek help using natural language queries with less syntactical precision \cite{Johnny}. However, \change{challenges persist in crafting effective prompts \cite{Johnny, Advait, Khurana_iui, Grounded_Copilot, Xu}, translating text-heavy outputs into actionable visual instructions, and in applying these LLM outputs within the software interfaces} \cite{Khurana_iui}. Emerging in-application LLM assistants or copilots \cite{cacm_article, copilot}, such as Microsoft 365 Copilot \cite{copilot}, Adobe Firefly \cite{Adobe_2023}, etc.) offer ways to automate software tasks based on prompts. \change{However, users face new challenges with these new LLM-based capabilities. Users have to navigate and build mental models of both the software's complex features \cite{Khurana_iui, Kiani_help} and the AI capabilities of the copilots, all while maintaining control in dynamic environments \cite{liao2023ai, Khurana_iui, Khurana_chatrex}}. \change{These hurdles have led to declining adoption rates for some copilots}, with many users choosing to entirely abandon them \cite{cacm_article}.

\begin{figure*}[!t]%
\centering
    {\includegraphics[width=0.95\linewidth]{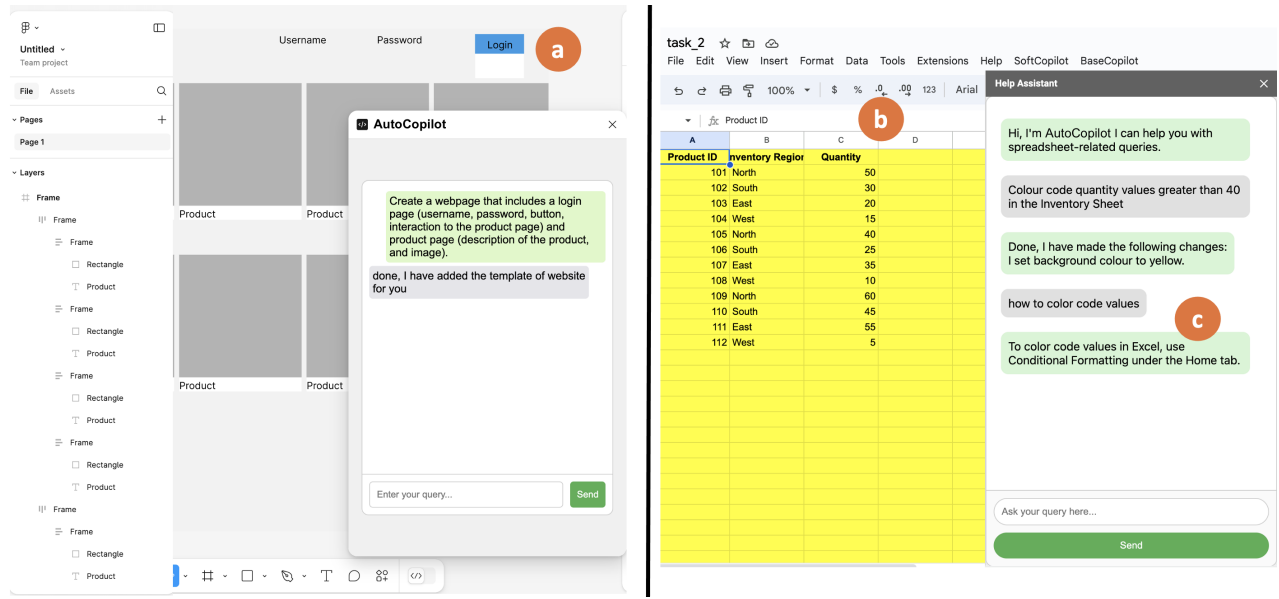} }%
       \vspace{-5pt}
    \caption{\fac{}: (a) Fully automates the user's task (e.g., creating a webpage that includes a login and product page); (b) Similar to state-of-the-art copilots, demonstrates incorrect automation (such as color coding the entire sheet instead of values greater than 40 in column C); (c) Provides follow-up textual response based on context from software documentation}%
   \label{FullAuto_ui}%
\end{figure*} 

Recent efforts to optimize copilots (e.g., for productivity or creative tasks) have largely focused on technical improvements, often without adequately considering user needs or incorporating insights from user studies \cite{tian2024spreadsheetllm}. Research shows that users prefer \textit{learning by doing} and resort to self-directed experimentation within an application \cite{carroll1984training, Novick, Rettig, Rieman, Kiani_help}. \change{The fully automated processes, while impressive, tend to bypass intermediate steps that prevent users from fully understanding specific software features and task-specific terminology} \cite{cacm_article, majeed_uist}.  \change{Recent literature also highlights concerns about balancing autonomy and agency, long debated in traditional AI systems \cite{endsley2017here, heer2019agency, shneiderman2020human}. \change{However, such concerns take on new dimensions in sophisticated LLM-powered copilots embedded in complex, feature-rich software tasks. In particular, the open-ended nature of interactions and the multifaceted landscape of user behaviors and lack of accurate user mental models poses unique challenges \cite{Khurana_iui}.} This creates a }new opportunity to provide more effective automatic and semi-automatic assistance. A critical question arises: do users prefer full automation or a balanced approach with semi-automation and guidance?  Our study contributes new insights into how users perceive \change{these two distinct design paradigms for} automation in copilots (semi-automation vs. full automation) when using feature-rich software.

\subsection{Human-AI interaction with LLM Assistants for Programming Tasks}
Although in-application software copilots have only recently started emerging, we can draw upon empirical studies of copilots for task-based assistance in domains such as programming and software development (e.g., GitHub Copilot within VS Code). While these copilots can automate code generation, they often require users to invest additional effort in reviewing and customizing the generated output to align with their actual intent and requirements \cite{ivie, businessinsiderCanceledMicrosoft, Grounded_Copilot, Vaithilingam}. These efforts exacerbate when the copilot generates incorrect automation (e.g., code generation) without explaining its actions, requiring users to invest time in understanding the generated output \cite{Grounded_Copilot, Vaithilingam, majeed_uist}. These findings suggest that users seek more than mere automation—they desire a deeper understanding of the syntax and steps involved in their tasks \cite{ivie}. 

Recent work has explored ways to better interpret complex, nuanced user inputs and generate more contextually relevant responses \cite{gao2020pile, prompt_catalogue, AI_2022}. However, despite spending time understanding LLM-generated code, users often find it harder to comprehend than their own code \cite{Naser, Grounded_Copilot, Liang, Mozannar,Xu}. As a result, many solutions focus on enhancing comprehension and explainability of the generated code to help users handle both familiar (where programmers already have some idea how to write the code) and complex programming tasks more effectively \cite{ivie, detienne2001software}. For example, the IVIE system supports in-situ understanding of code by augmenting the editor to help programmers grasp the generated content. Excessive automation can reduce user agency, while insufficient automation can make the assistant less effective or frustrating to use \cite{majeed_uist}. Our study extends these insights to the space of software copilots. Our main goal was to investigate how different levels of automated in-application help (semi-automation vs. full automation) impact task completion and user perceptions when using feature-rich software and we explored this by designing and implementing two interventions: \fac{} and \sac{}, described below. 

\begin{figure*}[!t]%
    \centering
    {\includegraphics[width=0.9\linewidth]{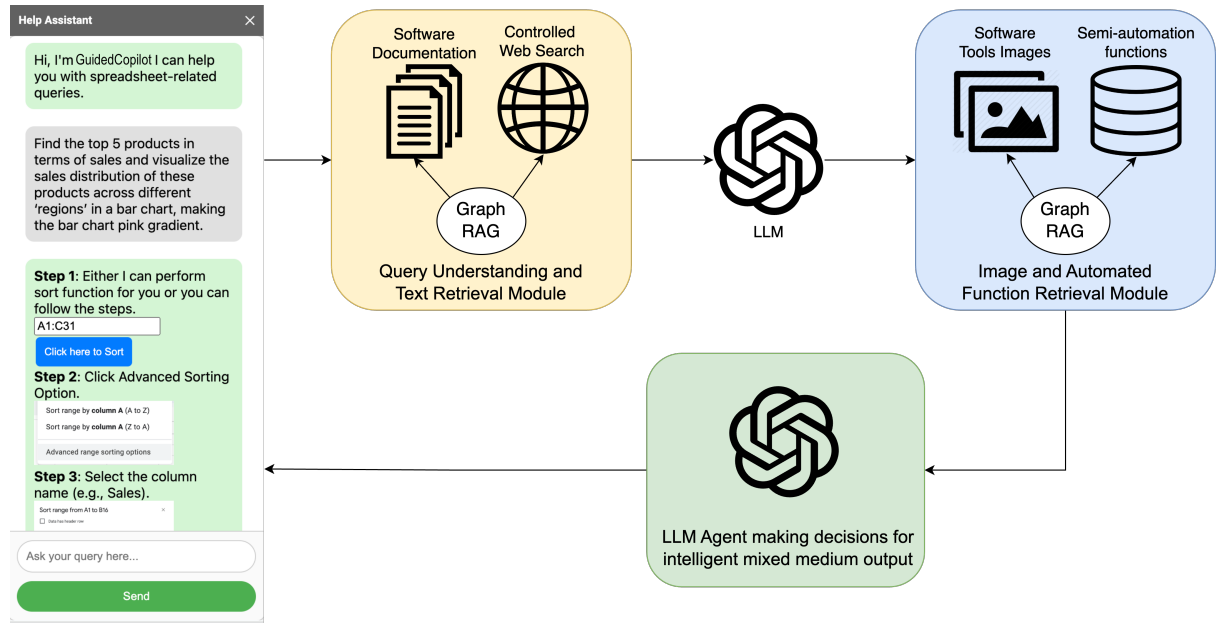} }%
      \vspace{-7pt}
    \caption{\sac{} Architecture: The user's query is used to initiate a conversation about automating software tasks, which is then transmitted to the query understanding and text retrieval module (Section \ref{text retrieval module}). This module interprets the query and performs a contextual search across documentation and web data. The extracted intent and relevant excerpts are processed by GPT-4o to generate text-based procedural steps. These steps are sent to the Image and Automated Function Retrieval Module (Section \ref{Function Retrieval Module}) for corresponding visual aids and automated functions, sourced from a curated dataset. Finally, the LLM agent integrates these text, visuals, and semi-automated functions into a cohesive, mixed-medium response tailored to the user's software-related query.}  %
   \label{SemiAuto_architecture}%
   \vspace{-9pt}
\end{figure*} 

\section{Intervention 1: Design Considerations and Implementation of \textit{\fac{}}}
\label{fullauto_designgoals}
\change{Our first copilot intervention, \fac{} (“\textit{Do It For Me}”), was based on full automation principles \cite{endsley2017here, Horvitz_2007, bainbridge1983ironies}} to serve as a \change{realistic} baseline (See Table \ref{table_comparison} for comparison of both copilots). \change{Such fully automated systems \cite{endsley2017here, Horvitz_2007, bainbridge1983ironies} are designed to maximize efficiency and task completion, often with limited or no human interaction.}  Drawing inspiration from \change{emerging automatic copilots} \cite{copilot, Adobe_2023}, \fac{} retrieves pertinent information from software documentation and web data to facilitate users in asking follow-up questions and presents a targeted and specific textual response to the user. \fac{} sometimes also demonstrates incorrect full automation, similar to modern copilots \cite{microsoftforum, microsoftForumsMicrosoft}. Based on these design considerations, we derived two key design goals for \fac{}:

\begin{enumerate}
    \item DG1: Provide full automation for the entire software task
    \item DG2: Offer concise follow-up responses based on context from the software documentation and web data, similar to in-application state-of-the-art copilots
\end{enumerate}

\subsection{\fac{}: User Interface Design and Implementation}
To maintain consistency and control in our experimental conditions, we chose to implement \fac{} ourselves to facilitate an easier data collection process. Our \fac{} UI (Figure \ref{FullAuto_ui}) enables users to input queries, which are then processed by GPT-4 to identify intent and extract relevant entities.  We created a dataset of basic automation functions tailored to the study's tasks, using AppScript in Google Sheets and JavaScript in Figma to control UI elements. We included instructions in the prompt for \fac{} to always attempt full automation for user queries. We implemented a hybrid-RAG using OpenAI’s text-embedding-3 model \cite{key} and Elasticsearch \cite{elasticHybridSearch, achiam2023gpt} to index the descriptions of these full automation functions and software documentation. These automation functions are then retrieved based on the user's intent and combined with extracted entities from the user’s query to generate the final output. The pertinent information from software documentation is used to generate a concise textual response and facilitate users’ follow-up debugging questions. \change{We integrated LLM-based web search for up-to-date follow-up responses when existing documentation fell short.} \change{To mirror real-world automation performance,} \fac{} \change {occasionally} encounters incorrect full automation when it fails to identify a single automation function from the multiple intents in the user's query for performing complex tasks outlined in the study, or when it cannot accurately map multiple entities from the user's query to the appropriate automation function. For example, in sorting data and applying color coding based on a condition, \fac{} may incorrectly map the entities to color code the entire datasheet rather than just the sorted values. Additionally, when the hybrid-RAG in \fac{} fails to identify relevant automation functions related to the detected intent, it shows a breakdown with an error message. \change{To ensure consistent reliability of automation across both copilot conditions, we employed the same software documentation for RAG, ensuring both copilots generated consistent response/automation quality and accuracy while minimizing hallucinations. Additionally, the researchers conducted sanity checks for both copilots to verify response accuracy.}

\section{Intervention 2: Design Considerations and Implementation of \textit{\sac{}}}

\label{semiauto_designgoals}
\change{We implemented \sac{} to reflect semi-automated systems that combine partial automation with user interaction and guidance to facilitate incremental learning \cite{endsley2017here, Horvitz_2007, bainbridge1983ironies, cacm_article}.} Drawing inspiration from software learnability research (discussed above), we considered how to structure the copilot assistance semi-automatically and how to include visuals in the step-by-step guidance. We developed four key design goals to enhance copilot assistance using visuals and semi-automation.

\textbf{Structuring the copilot assistance:} As copilot-driven task automation increases, crucial steps for users to learn software features are often skipped, reducing user control \cite{cacm_article}. When automation fails, these skills take control and manually perform these tasks. In such cases, \sac{} should provide guided help with visual references as this approach has been shown to be beneficial for feature-rich software tasks \cite{Khurana_chatrex, Kiani_help}. To balance user control and automation benefits, \sac{} should offer: (i) semi-automation for repetitive or trivial steps in the task, (ii) user control to trigger semi-automation when needed and, (iii)  human intervention to verify LLM-extracted data before automation proceeds. For example, simple tasks like “applying a sort filter in Google Sheets” or “creating a frame in Figma” could be automated, while detailed assistance could be provided for the more complex tasks.

\textbf{Enhancing the step-by-step guidance with visuals:} In-context GUI visuals \cite{Khurana_chatrex, Toby_multimodal, palmiter1991animated} have shown to be effective in software help-seeking. Since users often struggle to locate and use the correct UI elements when only text-based guidance is offered  \cite{brown_llm, AI_2022, copilot, Khurana_iui}, copilots provide relevant in-context visual references within the chat. These visuals, tailored to the user's specific task and application, can help users locate and use software features, such as the text tool in Figma or advanced sorting option in Google Sheets.

Based on the above considerations, we derived four key design goals for building \sac{}:

 \begin{enumerate}
     \item DG1: Provide step-by-step guidance with semi-automation for repetitive or trivial steps.
     \item DG2: Provide visual references to help users locate and apply the necessary software features.
     \item DG3: Allow users to initiate semi-automation and edit LLM-extracted entities before proceeding.
     \item DG4: Offer up-to-date mixed-medium support, combining web information with software documentation.
 \end{enumerate}

\begin{table*}[!t]
\centering
\begin{tabular}{p{2.4cm}p{6.2cm}p{5.5cm}}
\toprule
\textbf{} & \textbf{\sac{} \change{(“Do It With Me”)}: Semi-Automation Based Copilot \hspace{1cm} } & \textbf{\fac{} \change{(“Do It For Me”)}: Full Automation Based Copilot \hspace{1cm}} \\ \midrule
\textbf{Automation}
& Automate repetitive or trivial steps of the task (e.g., create a frame, shape, sort function, etc.), \change{initiated by users and} positioned at suitable steps within step-by-step guidance \hspace{10cm}

  \change{User can edit the LLM extracted entities from their query prior to performing semi-automation}
& Fully automate the entire software task (e.g, color code the quantities in column c greater than 40) \hspace{1cm} \\ \midrule
\textbf{Step-by-step \change{Visual} guidance} & Provides guidance on the task process.  \hspace{10cm}   

\change{Includes visual references in-context of a user’s tasks and application to help locate software features or functions and effectively apply them within the application}
& No \hspace{1cm} \\ \midrule
\textbf{Software Context and Follow-up questions }
& Up-to-date mixed-medium response by leveraging latest relevant information on the web along with software documentation \hspace{1cm}
& \change{Up-to-date} textual response \change{by leveraging latest relevant information on the web along with} software documentation \hspace{1cm}\\ \bottomrule
\vspace{-9pt}
\end{tabular}
\caption{Comparison of \change{two distinct design paradigms for automation:  \fac{} (“\textit{Do It For Me}”) vs. \sac{} (“Do It With Me”)}}
\vspace{-15pt}
\label{table_comparison}
\end{table*}

\subsection{\sac{}: User Interface Design and Implementation}
Based on the above design goals (Section \ref{semiauto_designgoals}), we designed and implemented \sac{} (Figure \ref{SemiAuto_ui}), a novel semi-automatic copilot that automates only trivial steps while providing \change{step-by-step visual guidance}. \change{Similar to \fac{},} we embedded \sac{} within Google Sheets and Figma to demonstrate its feasibility and scalability across diverse feature-rich software. Figure \ref{SemiAuto_architecture} illustrates the overall architecture of \sac{}, which can apply to any complex software with comprehensive software documentation, visual examples such as UI elements and screenshots, and an active user community on Q\&A platforms or forums. Additionally, the underlying software should support task automation through scriptable frameworks, making it suitable for a range of softwares, from productivity suites like Office 365 or Google Workspace, as well as creative tools such as Photoshop and AutoCAD. 

\subsubsection{Query Understanding and Text Retrieval Module:} \label{text retrieval module} 
When a user submits a query in the \sac{} UI, this module interprets the user's intent and retrieves contextually relevant information for generating accurate responses by implementing the state-of-the-art Graph-based Retrieval Augmented Generation (GraphRAG) \cite{githubGitHubMicrosoftgraphrag} using GPT-4o and the BAAI/bge-base-en embedding model \cite{bge_embedding}. GraphRAG uses NLP techniques and LLMs to construct dynamic knowledge graphs from documents, linking entities within and across sentences, outperforming traditional keyword searches and vector-based retrieval methods. To implement the GraphRAG, we used the following submodules: 

\textbf{Indexing:} We implemented LLM-based web searches to provide up-to-date responses for follow-up user queries when existing software documentation is insufficient. Using GPT-4o, we extracted search topics from user queries and conduct searches on targeted websites, such as Google Sheets and Figma Q\&A forums. The dynamically retrieved web data and documentation are segmented into discrete TextUnits or chunks, \change{from which entities (e.g., data items, software functions, GUI elements) and their relationships are extracted to construct the knowledge graph. The chunks are then} transformed into vectors and \change{indexed in a vector database for integration with GraphRAG}.

\textbf{Querying for fetching relevant data: } \label{Querying for fetching relevant data} After indexing, we performed querying on the index of software documentation and web data to extract relevant entities and paragraphs based on the user's query. These extracted entities and paragraphs were subsequently used as in-context information for prompts to GPT-4o to generate stepwise textual responses to the user's query, which were further utilized in Section \ref{Function Retrieval Module} to retrieve relevant images and semi-automation functions for integration into the appropriate steps.

\subsubsection{Image and Semi-automation Function Retrieval Module} \label{Function Retrieval Module}
This module consists of the following key components:

\textbf{Image and semi-automation function data preparation and indexing: } \label{Image and Semi-automation Function Data Preparation} We \change{scraped images from software documentation showcasing tool functionalities and UI elements, extracted their descriptions} and created an index using GraphRAG. We leveraged scripting capabilities in the form of plugins or extensions using HTML and JavaScript, often supported by feature-rich applications, to control UI elements and \change{create a set of semi-automation functions} (e.g., sorting data, creating charts in Sheets, or creating frames and buttons in Figma). We identified such 25 repetitive tasks for each application and created generic function templates capable of handling diverse user queries, which were then scripted to automate the tasks and indexed using GraphRAG.

\textbf{Querying for relevant images and semi-automation functions:} Finally, each text-based step generated by GPT-4o in response to the user's query (See Section \ref{Querying for fetching relevant data}) is treated as an individual GraphRAG query to retrieve relevant images and suitable semi-automation functions from the prepared indexes in Section \ref{Image and Semi-automation Function Data Preparation}. These retrieved images and semi-automation functions are then provided as input to the LLM agent.

\subsubsection{LLM Agent for Mixed Medium Output} \label{LLM Agent} We developed an LLM agent that decides how to strategically integrate the retrieved visuals and automation functions within LLM-generated textual step-by-step response. Following the design goals (DG1 and DG2)  as instructions, the LLM agent intelligently decides where to place these retrieved visuals and automation functions within the textual steps, producing a mixed-medium output presented to the user through the \sac{} UI. \change{Similar to \fac{}, \sac{} occasionally generated errors and exhibited uncertainty, when its GraphRAG fails to identify relevant semi-automation functions for the detected intent.}

Both the copilot UIs module was built as an in-application assistant and migrated to Chrome as an extension for Google Sheets and as a plugin for Figma. To allow users freedom and control in accessing the copilot UIs, we included the close options for closing it anytime during the interaction. The code repository and prompts for both copilot interventions are available upon request.

\section{Evaluation: Controlled Experiment and Follow-up Interviews}
To investigate the strengths and weaknesses of our two copilot interventions, we conducted a controlled experiment and follow-up interviews with 20 participants.

\subsection{Participants}
We recruited 20 participants (10F|10M) for our study, focusing on non-AI expert users with little to no prior experience or knowledge of ML or NLP; however, 4/20 participants had previous experience with ML or NLP. Our participants came from both CS (9/20) and non-CS (11/20) backgrounds (including Engineering, Accounting, History, Theory, Communication, and Arts) and professions (administrative, logistics, information designers, students, researchers, professional data scientist, and software developer). Participants were familiar with LLM-based assistants like ChatGPT (15/20), GitHub Copilot (7/20), and only 5/20 had used Microsoft 365 Copilot for software tasks before this study. About half of the participants (9/20) had frequently used Google Sheets and Excel applications and the remaining were occasional users. None of the participants had used Figma before. Our participants covered a range of age groups: 18-24 (32\%), 25-34 (47\%), 35-44 (16\%), 55-65 (5\%) and had different levels of education (1 Post-Secondary, 2 Diploma, 7 Bachelor’s, 6 Master’s, 4 PhD).  We recruited participants mainly from our university’s mailing lists and found additional participants through snowball sampling.

\subsection{Study Design and Procedure}
We used a within-subject design to reduce participant variability, employing a Latin Square counterbalancing \cite{raulin2019quasi} method to randomize the order of 2 copilot conditions (4 possible orders). Each participant completed two tasks per copilot (4 tasks in total), using one feature-rich application before transitioning to a second application, with the order of the tasks and copilots randomized.

At the beginning of a session, we introduced both copilots and provided tips for interacting with the application. Participants completed a demographic questionnaire on their background and prior experiences with different software applications, LLM assistants, copilots, and chatbots. Each copilot intervention was presented in random order, and tasks were designed to be complex enough to take at least 8 minutes, regardless of participant familiarity and prior experiences. After each task, participants completed a post-task questionnaire to evaluate their overall experience with copilot assistance, focusing on utility, user control, and potential for software learnability. We encouraged participants to think aloud \cite{norman2013design} and reminded them that the study was seeking to understand how they seek copilot assistance rather than their individual task performance. Lastly, we conducted follow-up interviews to probe into any difficulties that participants faced and their overall perceptions of different levels of automation. Each session lasted approximately one hour and participants received a \$15 Amazon gift card.

\begin{figure*}[!t]%
    \centering
    {\includegraphics[width=1\linewidth]{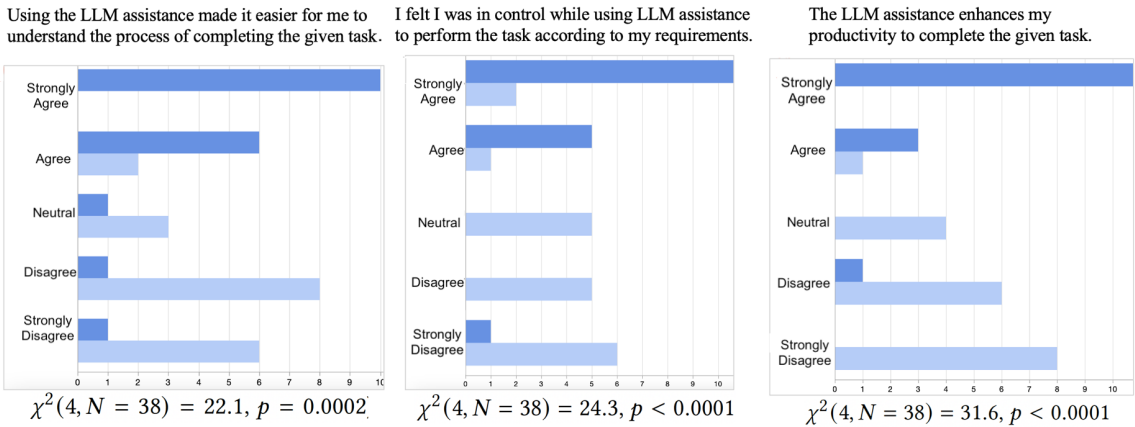} }%
      \vspace{-15pt}
    \caption{Overview of participants’ responses to the post-task questionnaire. The Pearson Chi-Squared test showed a significant difference for each metric across both copilot interventions for completing the Google Sheets and Figma tasks. With \sac{}, users demonstrated higher task completion and higher accuracy and indicated that \sac{} helped them learn the software-specific steps, provided users more control and enhanced their productivity compared to \fac{}. }  %
    \label{pearson_test}%
\end{figure*}

\subsection{Choice of Tasks and Applications}
We selected Google Sheets and Figma after exploring various other productivity and design-oriented applications (e.g., PowerPoint, Photoshop, Word), as they cover a range of tasks involving visual interactions, statistical functions, and interface design. To observe potential challenges copilot-assisted software help, we selected tasks that would require multi-stage help and prompts. For example, one of the tasks in Google Sheets was to use a copilot to analyze the top 5 products and visualize their sales across regions using advanced sorting and a custom bar chart. Similarly, in Figma, one of the tasks was to design a web page that includes a login section and displays products using copilot assistance. 

\subsection{Data Collection and Analysis}
We recorded each participant’s screen and audio recorded their interview responses. We focused on two key aspects: how they used copilot assistance and different levels of automation to complete the prescribed tasks in Google Sheets and Figma, and how they formulated and refined prompts to seek help. 

We ran Pearson’s Chi-square test for independence with between “Copilot Interventions” (having 2 levels: \sac{} and \fac{}) and user responses (having 5 levels, Strongly Agree to Strongly Disagree) to quantitatively determine the significance of the results. The experimenter, in consultation with all authors, compared task completion and accuracy against the ground truth, defined as the optimal sequence of steps and ideal application of software features \change{(See Appendix \ref{study_material} for details)}. To measure the task completion, we evaluated how many task steps (e.g., sequence of features/ functions) users completed using the copilots. To measure task accuracy, we assessed their success in correctly identifying and applying the sequence of features and functions (e.g., sort or VLOOKUP function, creating buttons or shapes).  \change{To study trial-and-error strategies, we manually annotated interaction logs to analyze repeated, varied attempts within user interactions, revealing non-linear task paths, repetitive actions, and corrective sequences \cite{campbell1960blind}.}

Finally, to complement our experimental findings, we corroborated the data with participants’ think-aloud verbalizations and probed into the reasons influencing users’ perception of utility, user control, potential for software learnability. We used an inductive analysis approach \cite{corbin1990grounded} and affinity diagrams \cite{corbin1990grounded} along with discussions among the research team to categorize the interview findings and identify key recurring themes.

\section{Results}

We have organized our results around key themes, evaluating each copilot's strengths and weaknesses in terms of utility, user control, productivity, and software learnability. We also report task completion rates, analyze user interactions to identify trial-and-error patterns, and highlight users' challenges and debugging strategies. 

\subsection{Task Completion, Software Utility and User Productivity with Copilot Interventions}

\begin{figure*}[!t]%
    \centering
    {\includegraphics[width=0.95\linewidth]{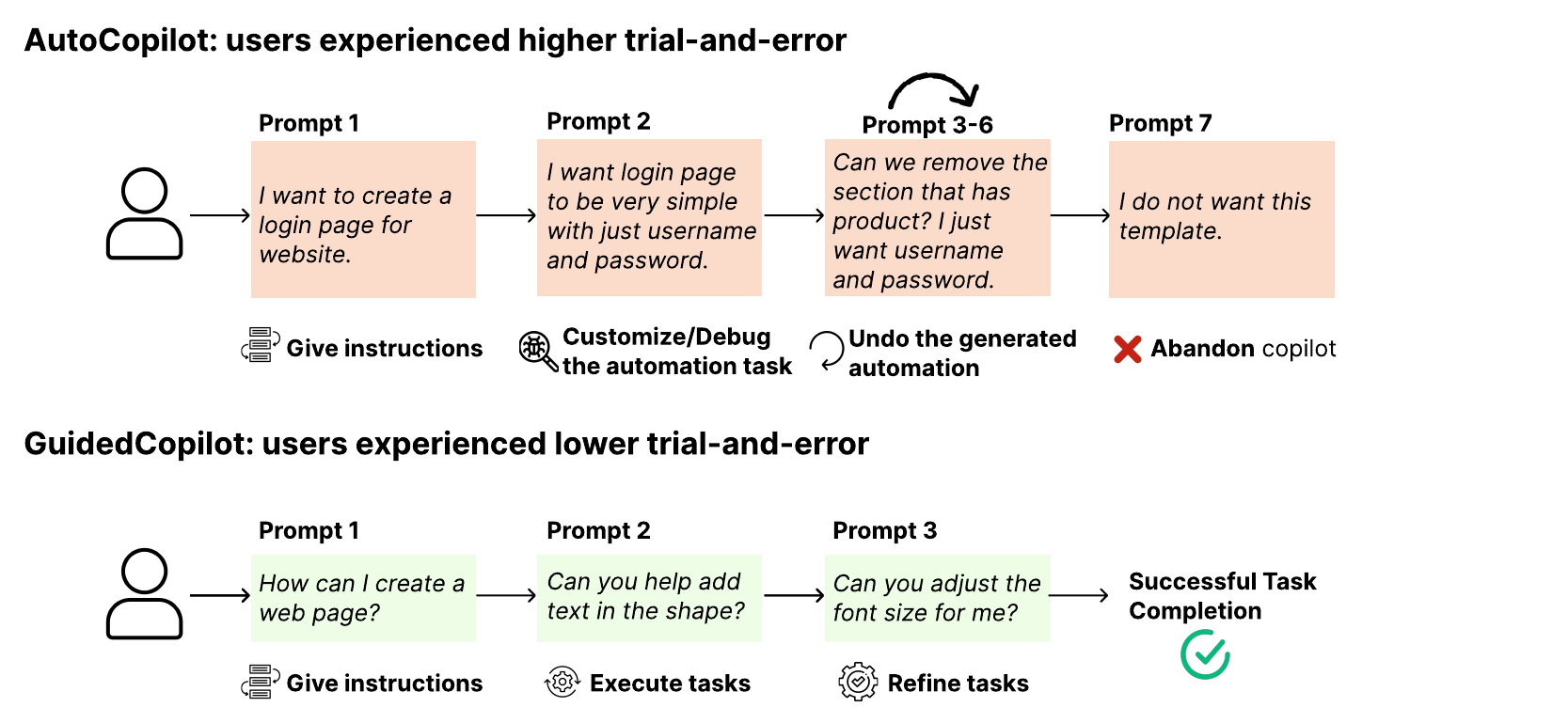} }%
    \caption{\textbf{Trial-and-Error Differences in  \fac{} vs \sac{}}: This figure illustrates the case of Participant P14, a computer science professional. Despite P14's technical expertise, they encountered higher trial-and-error with \fac{}, primarily focused on customizing and debugging the generated automation (e.g., a webpage template) and reversing (or undo) in case of incorrect automation, which ultimately led the user to abandon \fac{}. In contrast, P14 experienced fewer trial-and-error instances with \sac{}, mostly related to executing and refining tasks like resizing or altering colors. The user was able to complete the task successfully with \sac{}. }  %
   \label{case_study}%
\end{figure*}

\subsubsection{Task Completion and Task Accuracy: } 

None of the participants fully completed the tasks in Sheets or Figma using \fac{}. In contrast, with \sac{}, 11/20 participants successfully completed the tasks in Sheets, and 6/20 completed the Figma tasks, despite Figma being new to all users. On average, participants completed 35.0\% of the Sheets tasks (maximum= 50\% and minimum= 0\%) with \fac{}, compared to 88.5\% (maximum= 100\% and minimum= 70\%) with \sac{}. A paired-sample t-test further revealed a significant difference between the two copilots ($t(37.6) = 11.2$, $p < 0.0001$, two-tailed). For Figma, the task completion rate was 20.0\% (maximum= 50\% and minimum= 0\%) with \fac{} and 55.0\% (maximum= 75\% and minimum= 30\%) with \sac{}, also demonstrating a significant difference ($t(37.7) = 7.2$, $p < 0.0001$, two-tailed).

Among the portion of the task completed by each participant, task accuracy was significantly higher with \sac{} across both applications. In Sheets, the average task accuracy was 82.0\% (maximum= 100\% and minimum= 50\%) with \sac{}, compared to 12.0\% (maximum= 50\% and minimum= 0\%) with \fac{} ($t(37.9) = 17.2$, $p < 0.0001$). In Figma, the average task accuracy was 40.0\% (maximum= 60\% and minimum= 20\%) with \sac{}, and only 5.0\% (maximum= 30\% and minimum= 0\%) with \fac{} with significant difference ($t(30.7) = 10.4$, $p < 0.0001$) and no order effects were observed.

\subsubsection{User Perception of Copilot Interventions on Productivity and Software Utility:} Most participants (18/20) found that \sac{} enhanced their perception of productivity in completing Sheets tasks compared to \fac{}), and these results were significant ($\chi^ 2 (4, N=38) = 31.6$, $p<0.0001$), and align with task completion and accuracy scores. Even though Figma was new to all participants, most (14/20) felt that \sac{} enhanced their perception of productivity over \fac{} with significant differences ($\chi^2(4, N=38) = 18.9$, $p=0.0008$). Participants reported that \fac{} often reduced productivity by generating excessive or irrelevant automation, requiring substantial effort to adjust: \textit{“[Figma task]: I just need enough boilerplate to think through, but not too much. In this boilerplate, I had to spend extra time to change it...I felt like I was hitting my head against the wall...ends up wasting my time (P03).” } Conversely, 12/20 users appreciated that \sac{}'s user-initiated automation and visual instructions, which saved them time and enhanced productivity:\textit{ “I like the} \sac{}\textit{...the way it gives a mix of everything...text, images, and buttons...helpful and increases my productivity. It creates the frame, fields, like a built-in component interacting with the software...that’s step ahead (P11).”} 

Overall, most users preferred \sac{} over \fac{} across all key measures (See Figure \ref{pearson_test}). However, a few users (6/20), primarily male computer science professionals, favored \fac{} due to its complete automation and perceived time savings: \textit{“I would go with the complete automation...I am a professional and I don't care about how to make shapes in Figma. I just want to get the work done...it gives me a template I can improve and work with it (P17).”} These participants were comfortable debugging templates in Figma and felt it was easier than handling complex formulas and data manipulation in Sheets. An older participant (>55 years) also preferred \fac{} due to limited patience for manual experimentation: \textit{“As an older person… my patience is not high, I just want it to do it for me (P19).”}

\subsection{User Perception of Control with Copilot Interventions}
We next evaluated how participants felt about their ability to control and use the copilots : (i) while performing tasks, and (ii) when handling incorrect automation generated by copilots.

\subsubsection{In-Task Performance}

Most participants perceived greater control with \sac{} than with \fac{} when performing tasks in both Sheets (18/20) and Figma (16/20), with significant differences (Sheets: $\chi^2(4, N=38) = 24.3$, $p < 0.0001$; Figma: $\chi^2(4, N=38) = 15.7$, $p = 0.0035$). In contrast, with \fac{}, many participants (11/20) found full automation to be “a gamble”, especially in Figma which required extensive customization for visual tasks: \textit{“It felt like LLM 1} [\fac{}] \textit{was doing it for me,  which is a bit of a gamble...[as] I was little unsure of the result, whereas LLM2} [\sac{}] \textit{gave option of doing it with me, or here are the steps to do it myself (P04).” } With Sheets, many participants (13/20) also reported lower perceived control with \fac{}, as the lack of validation left them uncertain about the accuracy of complex tasks: \textit{“It} [\fac{}] \textit{felt like it had a mind of its own...it was doing whatever it wanted and all at once. I felt helpless and couldn't understand what went wrong. With data [tasks], things can get pretty bad (P12).”}

In contrast, \sac{} offered flexibility, letting users choose between step-by-step guidance or selectively using automation at suitable steps: \textit{“I would prefer} [\sac{}]\textit{...gave me more control to do it myself or get it done....not like a one-size-fits-all, and you can kind of fit or modify it to what you need it to do (P15).”} About half the participants (9/20) explicitly appreciated \sac{}'s detailed instructions that helped them verify the automation and reduce errors: \textit{“I feel more in control with step-by-step guidance. I can understand the process of that [task]...whereas If it did everything at once, I might not like a step in the middle...that did not get the results I was expecting and I'd have to undo (P08).” }

\subsubsection{Handling Incorrect Automation} 

When \fac{} produced incorrect automation, such as generating incorrect web page templates or coloring an entire spreadsheet instead of only quantities greater than 40 in a column, users often felt derailed and resorted to trial-and-error strategies. Participants engaged in 192 trial-and-error attempts with \fac{}, nearly double the 89 attempts with \sac{}. On average, users made five trial-and-error attempts per task with \fac{} (range: 3-10), compared to fewer than two with \sac{} (range: 1-5) across both applications. About 75\% of \fac{} attempts involved undoing incorrect automation, 55\% focused on further debugging or editing, and 10\% focused on locating and applying the assistance. These challenges led some participants (6/20) to abandon \fac{} entirely. 

In contrast, with \sac{}, \change{55\% of 89 trial-and-error efforts} focused on enhancing or customizing tasks (e.g., resizing shapes or altering text colors), with only 15\% focused on correct the automation. For example, \change{P14, faced} numerous debugging challenges (Figure \ref{case_study}) with \fac{}’s webpage template. However, with \sac{}, they only made minor adjustments, such as resizing or altering colors: \textit{“}\fac{} \textit{led me into a spiral of figuring out what to do next...I just wanted a simple page with a username/password, but it kept adding things back and [would] not let me change certain things I wanted. I felt lost and caught in a spiral of bad [prompting] decisions (P14).”}

\subsection{User Perception of Potential for Software Learnability}

Most participants (16/20) indicated that \sac{} improved their ability to learn the steps of the software tasks compared to \fac{}, with results significant for both Sheets ($\chi^2(4, N=38) = 22.1$, $p = 0.0002$) and Figma ($\chi^2(4, N=38) = 18.9$, $p = 0.0008$). Participants found \sac{}'s in-context visual cues particularly helpful for locating relevant menu items and navigating applications like Sheets: \textit{“The best thing about this LLM} [\sac{}] \textit{is it showed images as a visual clue with instructions. I just had to match those little pictures with the icons in the sheets, sort of teaching and explaining the process which helped me learn how to use the software and execute the task (P16).”}

\begin{figure*}[!t]%
    \centering
    {\includegraphics[width=0.95\linewidth]{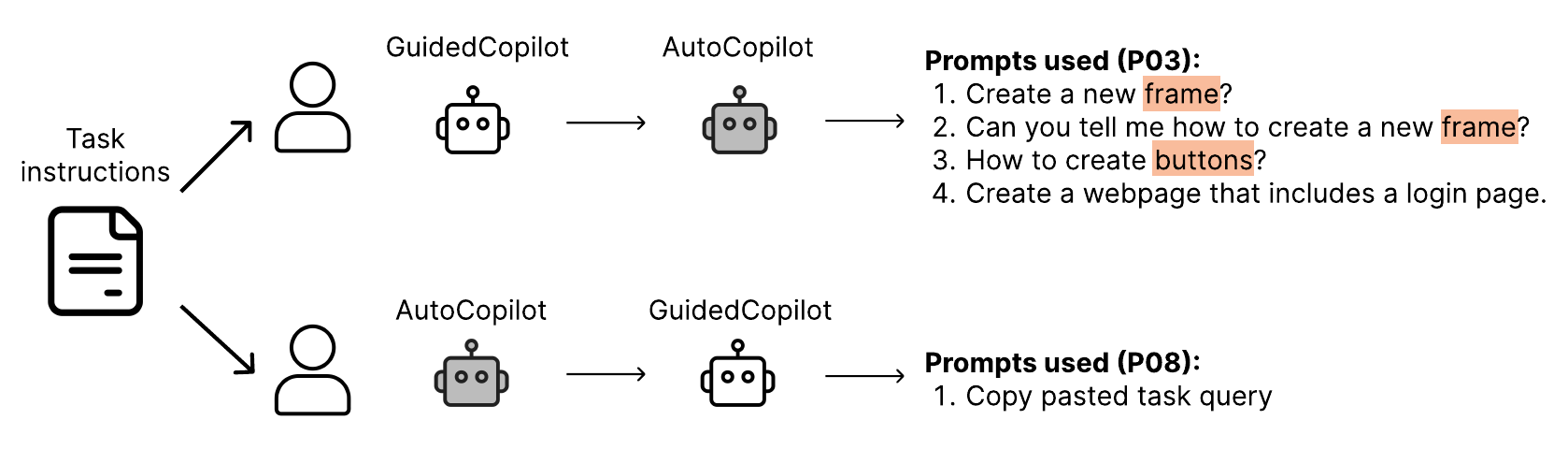} }%
      \vspace{-10pt}
    \caption{Illustration of Transfer Learning in Software Usage with \sac{}: Participant P03 started with the \sac{} condition where the step-by-step instructions helped them grasp fundamental software features (e.g., creating frames and buttons in Figma), which they effectively transferred when doing subsequent tasks. In contrast, P08, who started with \fac{}, did not demonstrate similar foundational knowledge and continued to rely on a repetitive prompting strategy of copy-pasting queries in follow-up tasks.}  %
   \label{transfer_learning}%
    \vspace{-5pt}
\end{figure*}

For unfamiliar applications like Figma, \sac{} helped users learn basic software features, with half transferring their knowledge to new subsequent tasks (known as transfer learning, (See Figure \ref{transfer_learning}). As P08 commented: \textit{“}\sac{} \textit{helped me like a tutor, giving visuals and instructions that take you easily through the interface and helped me learn and gain confidence in using the software, so I learned how to create a frame, use the rectangle tool (P08).”} In contrast, many users (11/20) felt dependent on \change{\fac{}, expressing doubt about whether }they could perform similar tasks independently in the future: \textit{“I don’t want to rely entirely on the AI} [\fac{}] \textit{to do everything; I want to learn the software while doing the task. AI should help us learn to complete tasks...not make us feel dependent on it (P05).”}

Overall, most participants (15/20) were enthusiastic about using \sac{} for future software tasks: \textit{“I like} \sac{} \textit{because it saves time for my accounting tasks to check on payments and offers both options-doing things automatically and showing how to do them myself (P16).”} Participants also expressed interest in using \sac{} for other complex applications like Photoshop and SolidWorks.

\subsection{Areas of improvement in \sac{}}

Despite the overall positive interaction with \sac{}, a few participants faced minor usability issues, such as difficulty in mapping visuals within the chat to the software’s UI: \textit{“I could not...find where to locate the VLOOKUP [in Sheets]; maybe highlighting a section of the page with 'click here' would be more helpful (P11).”} A few participants (5/20) struggled with the suggested steps, especially when new to the software. Participants requested more animated instructions tailored to the real-time context of the task progress and the application’s status: \textit{“I could not use the rectangle tool...having an instructional tutorial that better understands the tasks and recognizes that I clicked on the menu bar with rectangle option and then give me instructions to learn how to use it would be helpful” (P19).}  Finally, some advanced users wanted even \textit{more} automation and fewer steps aligned with their current task progress.
 
\section{Enhancing User Experience with Semi-Automatic Copilots: A Follow-up Design Exploration}

\begin{figure*}[!t]%
    \centering
    {\includegraphics[width=0.9\linewidth]{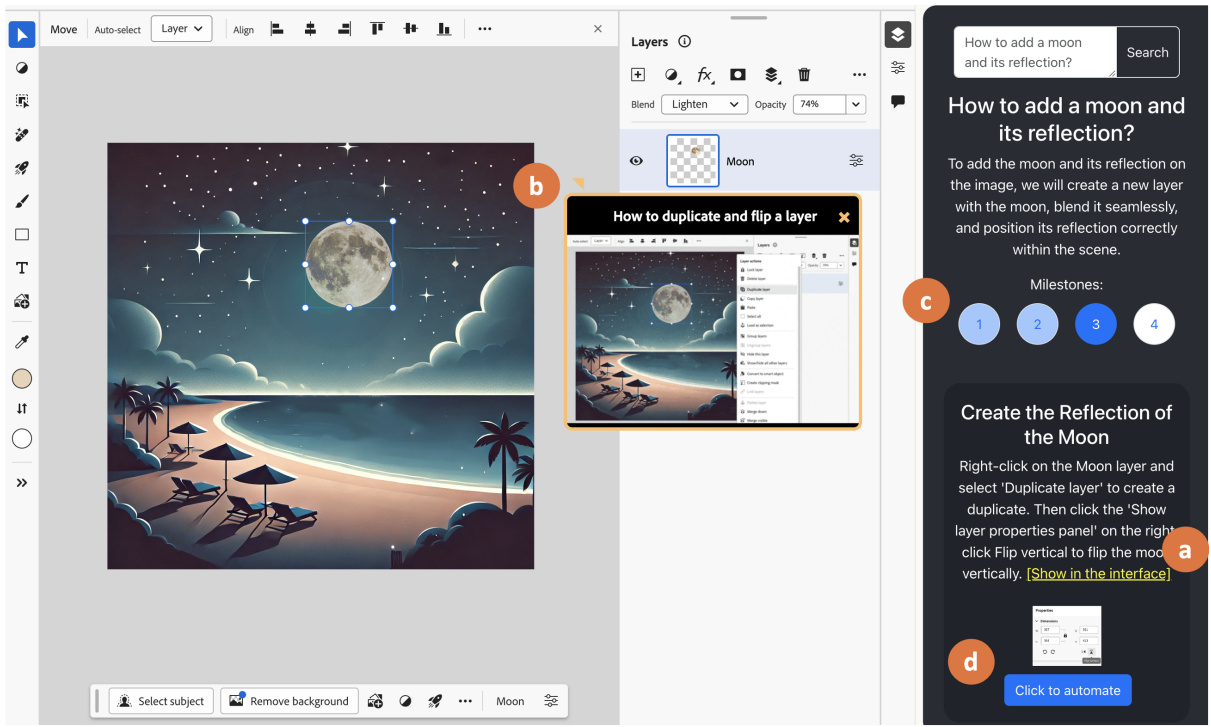} }%
      \vspace{-2pt}
    \caption{\textbf{Design exploration} building on \sac{} to offer targeted, task- and state-aware assistance in Photoshop through two key features: (a, b) \sav{} integrates in-context preview clips within the software interface. When the user clicks “Show in the interface”, the copilot detects the user’s task progress and application state to display relevant preview clips for the next step (e.g., after adding and resizing the moon, the clip suggests the menu option for creating a reflection). (c) \saa{} adapts the number of instructions based on user needs; for example, if the user has already adjusted the moon's size, it skips to step 3. Clickable milestones provide control and flexibility to navigate through instructions. (d) The user can opt for semi-automation by clicking on relevant options.}  %
   \label{SemiAuto_vst}%
    \vspace{-5pt}
\end{figure*}

Based on the insights from our first study, we identified key design considerations for further enhancing user experience with semi-automatic copilots that were rated as being more useful overall. We explored variations for semi-automatic copilots that build on \sac{} to better integrate visuals and can provide more targeted, task- and state-aware assistance (See Figure \ref{SemiAuto_vst}). For this exploration, we selected Adobe Photoshop, as users had frequently expressed a desire to have access to semi-automatic copilots in this feature-rich application. To simulate both of these adaptive approaches, we employed the Wizard-of-Oz method \cite{Wizards_Porcheron, Dahlb}. To capture the users' initial reaction to these designs, we conducted a usability study with 10 participants who had varying experiences with the software application.

\subsection{Feature 1: \sav{} (Visual Step-through)} \sav{} (Figure \ref{SemiAuto_vst}b) offers dynamic step-by-step assistance using visual anchors and context-sensitive preview clips specific to the user's current tasks and the application state. Our design complements prior research that has shown the value of in-application instructional videos \cite{fourney2012then, kelleher2005stencils, alliefraser} and step-through demonstrations \cite{Khurana_chatrex, Lafreniere_Tutorials, Novick_tutorials, Grossman_ToolClips} by incorporating users' task progress and application state. When users select “Show in the interface” (Figure \ref{SemiAuto_vst}a), \sav{} presents targeted preview clips that dynamically adapt to the user's progress and application state, visually demonstrating the next steps (e.g., after adding and resizing a moon, it suggests options for creating a reflection). To design this feature, we prepared the preview clips relevant to \change{each step} in advance, along with relevant images and screenshots, to support users in locating the corresponding menu items within the application.

\subsection{Feature 2: \saa{} (Adaptive Mixed-medium)} \saa{} provides in-context mixed-medium assistance tailored to the user's current task and application state, dynamically adapting instructions based on the user's needs (Figure \ref{SemiAuto_vst}c). For example, if the user has already adjusted an element, \saa{} skips to the next relevant step (e.g., step 3). We used GPT-4o to generate textual step-by-step instructions. The UI module is built using ReactJS and migrated to Chrome as an extension. The researcher monitored real-time task progress and the application's state and manually controlled which steps needed to be shown next, facilitating rapid prototyping and validation of various adaptive features while gaining  insights into user interactions. The researcher played the preview clip based on the user's completed tasks and the next required steps, effectively mimicking a task- and state-aware adaptive system. To simulate semi-automation, the researcher executed these actions on behalf of the system whenever the user opted to automate a particular step.

\subsection{Study Procedure and Participants}

To study user’s initial reactions, we conducted a usability study with 10 participants (4F|6M) aged 20-30, from \change{different} backgrounds (CS, Robotics, Engineering) and education levels (5 Master’s, 3 PhD, 2 PostDoc). About half of the participants (6/10) were novices, while 4/10 were experienced Photoshop users. All participants were familiar with LLM-based assistants like ChatGPT, and some with GitHub Copilot (4/10).

Each session began with an introduction to both features within the copilot variation, followed by a demographic questionnaire on participants' backgrounds and prior experiences with LLM assistants and software applications, lasting 30 minutes in total. They were then asked to complete tasks using the web version of Adobe Photoshop with the \sac{} plugin installed. The task involved adding a moon image and its reflection to a scene and integrating them seamlessly with the background (Figure \ref{SemiAuto_vst}), using both in-application copilot features (\sav{} and \saa{}) for assistance. In follow-up interviews, we probed further into user reactions and overall experience. Participants were encouraged to think aloud throughout the session. Sessions were video and audio-recorded for transcription, including the actions on screen. 

\subsection{Findings}

\subsubsection{Users' reactions when using \saa{}}
All expert users (4/10) appreciated \saa{} for skipping the steps they had already completed, especially in tasks where they were proficient and had clearer goals: \textit{“The milestones are structured...it directs me from step one to three when I already completed step two. It works when you have a clear goal and you know there’s one way to go...(D04).”} They suggested even more granular detection of actions to assist with option-based scenarios (e.g., opacity and blend). Experts also saw the potential for this feature in apps like Canva, Cadence, and SolidWorks. 

Novices appreciated the clickable navigation of milestone steps for unfamiliar tasks and the ability to skip those steps when proficient. They found this adaptability helpful for validating the successful completion of their current step: \textit{“I like it} [\saa{}] \textit{...it breaks down steps and automatically goes to the next step when I finish...I don't have to check if I completed that step. If I want to explore the other functions I can go back (D06).”} Both novice and expert users noted that \saa{} would be increasingly useful as novices gained proficiency with the application, for both complex and artistic tasks.

\subsubsection{Users' reactions when using \sav{}}

All participants found the task and state-aware preview clips easy to use and helpful for forming a mental model of the user interface: \textit{ “I know what is going on in my mind...is clear with these videos [preview clips] after reading the steps, it directly shows me which button to click, which parameter to modify (D04).”} Unlike traditional video demonstrations, participants liked that these clips  aligned with the application's current state and task progress, saving them time and helping locate UI elements more quickly: \textit{“I like preview clips based on my task progress because they save my time...I don't need to go to [the] beginning every time...making it more automatic was helpful (D09).”} Participants were eager to see these \change{clips for tasks} in other feature-rich applications like SolidWorks, Canva, and Lightroom.


\section{Discussion}
\subsection{Key Takeaways} In this paper, we provide \change{compelling evidence that users value maintaining control when interacting with software copilots, particularly for complex tasks.} \change{Semi-automation, as implemented in \sac{}, was effective for repetitive and trivial steps, enhancing perceptions of productivity and user satisfaction. While \fac{}'s full automation appealed to a few of the technical participants, it was often misaligned with tasks and increased users' debugging efforts}. In contrast, \sac{}’s step-by-step \change{visual guidance empowered users to learn software skills, validate automated processes, and regain control when the automation fell short}, especially for unfamiliar tasks. \change{While our study focused on \sac{}'s implementation} in two applications, our implementation using scriptable frameworks is flexible and can be generalized to other feature-rich software with detailed documentation and visual examples. Furthermore, our follow-up design exploration \change{highlights opportunities to enhance} semi-automatic copilots with task- and state-aware \change{features that dynamically adapt to user needs}. 

\begin{figure*}[!t]%
    \centering
    {\includegraphics[width=1\linewidth]{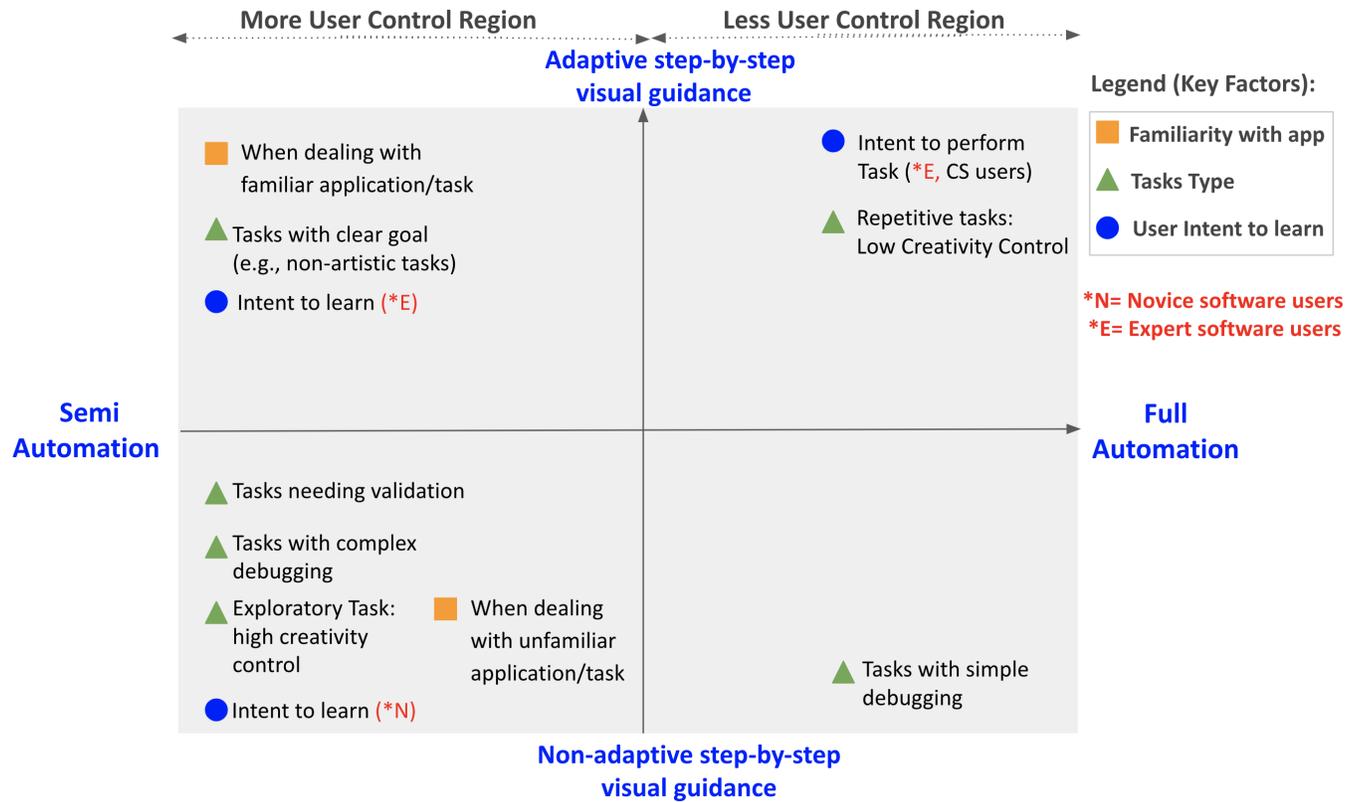} }%
      \vspace{-15pt}
    \caption{Dimensional framework describing key factors to consider when determining levels of automation and step-by-step guidance in copilots: (a) \textbf{Familiarity with the application:} When designing copilots for unfamiliar applications, a guided, semi-automatic approach with visual references can help users onboard, while more adaptive support for experts balances time savings with user control, catering to different expertise levels. (b)\textbf{ Tasks Type: } Higher levels of automation are best suited for straightforward, repetitive, or simpler visual tasks. For more nuanced tasks involving complex decision-making, debugging, user dependencies, or creative input, semi-automatic copilots with \change{step-by-step visuals or previews} are more beneficial, allowing for greater user control. More adaptive guidance is useful for tasks with clear goals and intentions, typically non-artistic and non-exploratory, within feature-rich applications. (c) \textbf{User Intent to Learn:} For users with a clear intent to learn while using feature-rich software, future copilots should adopt a semi-automatic approach with step-by-step guidance—more adaptive for experts and less adaptive for novices—balancing automation with user learning. For those focused on time efficiency, such as expert software users and CS male users, a higher level of automation would be more suitable.}  %
  \label{dimesnion}%
\end{figure*}

\change{By challenging the trend toward fully automated copilots, our findings underscore the importance of end-user customization, guidance, and control for effective user-copilot collaboration.} Our \change{experimental and qualitative insights contribute to a deeper understanding of how to design copilots for complex workflows within feature-rich applications that} accommodate different experience levels. \change{Consistent with historical lessons in automation research \cite{cacm_article}, we argue that AI and feature-rich interfaces will coexist and evolve together as complementary tools rather than replacements}. HCI and AI researchers, interface designers, developers, and others working on LLM-powered assistants \change{leverage our framework and key factors (Figure \ref{dimesnion}, Section \ref{dimension_framework}) to achieve more effective human-AI interaction.}

\subsection{Key factors to consider for levels of automation}
Based on our two study findings, we identify three key factors for determining the appropriate level of automation and user control in software copilots. We propose a three-dimensional framework (Figure \ref{dimesnion}) with axes for Semi-Automation vs. Full Automation, Adaptive Step-by-Step Visual Guidance vs. Non-Adaptive Step-by-Step Visual Guidance, and three influencing factors: (i) Familiarity with the application, (ii) Task Type, (iii) User Intent in learning vs. performing tasks.

 \label{dimension_framework}
\subsubsection{Familiarity with the application}$\tikz \fill[orange] (0,0) rectangle (0.2,0.2);$ Our findings highlight that users benefit more from semi-automatic copilots with \change{stepwise visual guidance}, particularly when onboarding with unfamiliar software. While full automation may streamline trivial tasks, it often lacks explainability and does not help users develop a mental model of software features or learn how to manipulate automation. Future research should explore how software copilots shape user mental models, especially for novices. Though some studies have explored mental models with LLM assistants in help seeking \cite{Khurana_iui, Subramonyam}, more research focused on the new generation of in-application copilots is needed \change{to foster} more intuitive and effective interactions. \change{Future work can also explore long-term deployments to investigate the impact of interaction duration (short-term vs. long-term) on user expectations and experience with automation in software copilots.}

As users gain experience with a software, adaptive semi-automatic support has the potential to enhance software utility and productivity. Building on prior research on personalized and adaptive help systems \cite{Delisle, augstein2023personalized, Pangoli, Hastie, Matejka_CommunityCommands}, future copilots should be designed to offer adaptive, semi-automated approaches that balance time savings with user control, catering to varying expertise levels.

\definecolor{grassgreen}{RGB}{124, 252, 0}
\definecolor{redpink}{RGB}{255, 102, 153}
\subsubsection{Task Type} $\tikz \fill[grassgreen] (0,0) -- (0.15,0.3) -- (0.3,0) -- cycle;$ \label{tasks_type} Our study shows that higher levels of automation work best for simple, repetitive tasks aligning with prior works \cite{cacm_article}, \change{but semi-automation is preferred} for complex decision-making, debugging, or validation tasks (e.g., intricate data analysis or exploratory design).  Copilots that offer adaptive visual guidance and previews can be particularly helpful for \change{creative software tasks. Such copilot assistance} can prevent users from becoming passive monitors — a risk often transformed with higher levels of automation \cite{cacm_article}. Future copilots should leverage both task categories and software, moving beyond the general-purpose designs \change{(e.g., \cite{copilot, figma})} to better cater to diverse user needs. The HCI and AI communities should join forces to focus on creating copilots tailored to users' specific tasks and needs.

\subsubsection{User intent in learning vs. performing tasks}$\tikz \fill[blue] (0,0) circle (0.15);$ Although we did not formally investigate gender, age, or CS background as a factor, we observed that male users with CS backgrounds preferred automation for time efficiency. However, fully automated approaches risk overreliance and deskilling \cite{cacm_article, Khurana_iui}, particularly for novices. Users can face a diminished ability to perform tasks independently or intervene when automation fails. Most novices preferred our semi-automated copilot that promoted learning \change{of software steps}. This aligns with established help-seeking behaviors, such as “learning by demonstration”, using visual mediums \cite{Toby_multimodal, Khurana_chatrex, palmiter1991animated}, step-by-step guidance \cite{Chilana_lemonaid, Khurana_chatrex}, adaptive help \cite{Delisle, augstein2023personalized, Pangoli, Hastie, Matejka_CommunityCommands}, in-context help systems \cite{Brandt_joel, Chilana_lemonaid, Grossman_ToolClips, Hartmann, Lafreniere_incontext}, and visual demonstrations \cite{Lafreniere_Tutorials, Novick_tutorials, Chilana_lemonaid, fourney2012then, kelleher2005stencils, alliefraser}, all proven effective for enhancing user learning and retention. \change{Future semi-automatic copilots could go a step further by integrating Explainable AI (XAI) \cite{Liao_XAI, miller2019explanation} to make AI decision-making processes more understandable. Semi-automatic copilots could not only assist in task execution (as seen in \sac{}) but also help users learn from the AI's reasoning \cite{Kim_XAI, Khurana_chatrex}, ultimately fostering skill development and reducing overreliance. Such} copilots for complex software tasks should balance automation with learning, incorporating mixed-medium, in-context help, as seen with \sac{}. Our paper provides empirical evidence that copilots should function truly as “co-pilots” \cite{cacm_article}, supporting users without diminishing their role or skills, and draw from established help-seeking strategies \cite{carroll1984training, fourney2012then, kelleher2005stencils, alliefraser} to maintain a balance between automation and user learning.  

\subsection{Prioritizing User Control When Designing Copilots}

As seen in studies of LLM-powered tools for code generation \cite{majeed_uist, cacm_article}, our findings \change{show that users preferred \sac{} as its balanced automation with instructional} guidance allowed users to choose their level of control. \change{Our follow-up study further suggests that} while automating certain tasks can enhance efficiency (See Section \ref{tasks_type}), users need the ability to intervene, make decisions, understand the process, and retain control. This is important in particular when dealing with complex or creative tasks where automation has higher chances of failing. This approach aligns with the concept of mixed-initiative interactions \cite{Allen, Horvitz_2007}, where users shift between manual and automated modes based on their needs and task complexity. 

Although we did not focus on creativity needs, previous research suggests that visual artists and graphic designers also value control over their creative processes when using visual \change{feature-rich software} \cite{HyungKwon, Jahanlou}. Future research should explore how copilots can better support creative workflows by offering adaptable automation that enhances, rather than undermines, their creative autonomy. By allowing users to select their preferred level of automation, future copilots can better support diverse user and task needs, fostering both effective collaboration and continuous learning. Future research should also explore multi-agent strategies to optimize when and how automation should intervene, ensuring that copilots empower users while maximizing the benefits of automation. 

\section{Limitations}

Our study \change{examined two distinct design paradigms} in software copilots (semi-automation vs. full automation) \change{and how they} impact task completion and user perceptions. Although our findings emphasize the importance of user control and accommodating diverse experience levels, they are limited by the applications used (Figma, Sheets) and the evolving capabilities of LLMs. Although our prototypes were effective for initial insights, advanced vision-based LLMs like SORA could provide more real-time data and deeper analysis. We did not directly assess the impact of specific features (e.g., automation, step-by-step guidance), as this would compromise ecological validity. Our findings highlight the issue of full automation without transparency or guidance, suggesting that automation alone is insufficient. Future research should explore the role of transparency in both automation paradigms and quantitatively assess the independent effects of these features, along with individual differences (e.g., age, gender, expertise), using larger, more diverse samples.


\section{Conclusion}

We investigated \change{two design paradigms for automation in software copilots}: \fac{}, a fully automated copilot, and \sac{} that combines automation of trivial steps with step-by-step visual guidance. Our results show that while full automation may appeal to a few users, most prefer to maintain control over complex tasks, favoring semi-automation for repetitive and trivial steps. As copilots advance toward full automation, our findings underscore the importance of offering greater user control and accommodating diverse experience levels to maximize effectiveness. These insights provide valuable guidance for HCI and AI researchers, designers, and developers in balancing automation with control and user autonomy, fostering more effective human-AI collaboration in feature-rich software.

\begin{acks}
We thank the Natural Sciences and Engineering Research Council of Canada (NSERC) for funding this research.
\end{acks}
\bibliographystyle{ACM-Reference-Format}
\bibliography{sample-base} 
\newpage
\appendix
\clearpage
\section{Additional details on Data Analysis: Controlled Experiment and Follow-up Interviews}
\label{study_material}
\change{We describe the actual software tasks users performed with assistance from copilot interventions, along with the ground truth for each task, to evaluate task completion and accuracy.}

\subsection{\change{Ground truth sample for spreadsheet tasks in Sheets}}

\begin{itemize}
  \item \change{\textbf{Task 1:} Participants were asked to use copilot assistance to find the top 5 products in terms of sales and visualize the sales distribution of these products across different ‘regions’ in a bar chart, making the bar chart pink gradient.\\
    \textbf{Ground Truth: } The ground truth for successful task completion is defined by how many of these task steps users completed (1) using the sort function on the sheet with the data ranging from A1:C31 (2) creating a bar chart of the data ranging from A2:C6 (3) using the customize tab in the chart editor to modify the Fill color and Line color of the series bar to pink in the bar chart. The ground truth of the task accuracy reflects how accurately they applied these outlined sequence of steps and software functions to complete the given task. 
     \item \textbf{Task 2:} Participants were asked to use copilot assistance to color-code quantity values greater than 40 in the Inventory sheet and to find and insert the corresponding product names from the 'Products' sheet into column D of the ‘Inventory’ sheet. They were asked to match the product IDs listed in column A of the ‘Inventory’ sheet with those in the 'Products' sheet under the 'Products' column.\\
    \textbf{Ground Truth: } The ground truth for successful task completion is defined by how many of these task steps users completed (1) using the Conditional Formatting on the selected data range C2:C13 (2) applying the “Greater than” format rule to set the condition for values greater than 40 (2) using the VLOOKUP formula (e.g., =VLOOKUP(A2, Products!A:B, 2, FALSE)) in column D of the ‘Inventory’ sheet, and (3) copying the formula down column D to apply it for all product IDs. The ground truth of the task accuracy reflects how accurately they applied these outlined sequence of steps and software functions to complete the given task.} 
\end{itemize}

\subsection{\change{Ground truth sample for UI design tasks in Figma}}

\begin{itemize}
  \item  \change{ \textbf{Task 1:} Participants were asked to use copilot assistance to create a webpage that includes a login page (username, password, button, interaction to the product page) and product page (description of the product, and image). (Participants were provided with the reference webpage to create)\\
    \textbf{Ground Truth: } The ground truth for successful task completion is defined by how many of these task steps users completed (1) using the Frame (F) tool to create two frames for login and product page function on the sheet (2) using the Rectangle tool to add input fields for username and password and create Submit button (3) using the Text Tool (T) to add a label above each field (e.g., “Username” , “Password”, “Submit”) and adjusting the spacing of the added rectangles (4) using Image tool to include prodcut images (already provided to the participants) (5) adding interactions to the button using the Prototype tab and dragging prototyping handle from login button to product page. The ground truth of the task accuracy reflects how accurately they applied these outlined sequence of steps and software functions to complete the given task.   
     \item \textbf{Task 2:} Participants were asked to use copilot assistance to design an interactive tutorial that provides a comprehensive three-step interactive project timeline of your product development cycle and looks like the reference project timeline. (Participants were provided with the reference project timeline) \\
    \textbf{Ground Truth:} (1) using the Frame (F) tool to create a frame (2) using the Eclipse tool to draw an oval shape for the timeline, Rectangle tool for creating buttons and adjusting the size, spacing and shape (3) using the Fill option to color the shapes and Text Tool (T) to add a label above each field (e.g., “Step 1: Brainstorming” , “Step 2: Prototyping”, “Next”) (4) adding interactions to the button using the Prototype tab and dragging prototyping handle from Next button to second page The ground truth of the task accuracy reflects how accurately they applied these outlined sequence of steps and software functions to complete the given task.}
\end{itemize}
\end{document}